\RequirePackage[2020-02-02]{latexrelease}
\documentclass[aps,prd,groupedaddress,graphicx,nofootinbib]{revtex4}
\usepackage{amsmath,amssymb,graphics,graphicx,color,epsf}
\usepackage{subfigure}
\usepackage[scanall]{psfrag}  
\usepackage{tikz}

\begin{document}

\title{Hilltop Sneutrino Hybrid Inflation}

\author{Chia-Min Lin}

\affiliation{Fundamental General Education Center, National Chin-Yi University of Technology, Taichung 41170, Taiwan}



\begin{abstract}

In this work, we consider a hilltop version of the (supersymmetric) sneutrino hybrid inflation where the right-handed sneutrino field plays the role of the inflaton field. This model is a type III hilltop inflation that can produce a spectral index $n_s=0.96$ which fits perfectly to experimental observations without fine-tuning of parameters. We also briefly consider nonthermal leptogenesis via the decay of the right-handed sneutrino inflaton field after inflation.  

\end{abstract}
\maketitle
\large
\baselineskip 18pt
\section{Introduction}

Inflation \cite{Starobinsky:1980te, Guth:1980zm, Linde:1981mu} is arguably the most popular scenario for the very early universe cosmology. Although experimental observations (such as Planck \cite{Planck:2018jri}) have already ruled out a countless number of models, there is still no consensus about which inflation model is the best and what is the inflaton field. Roughly speaking, inflation models can be divided into two categories. There are large-field models in which the inflaton field value is larger than the reduced Planck mass $M_P \equiv (8\pi G)^{-1/2} \simeq 2.4 \times 10^{18}\mbox{ GeV}$ when our universe is leaving the horizon. On the other hand, there are small-field models in which the corresponding inflaton field value is smaller than $M_P$. A characteristic feature of large field inflation models is the prediction of stronger signals of primordial gravitational waves originating from the quantum fluctuations of the tensor perturbations. Experimentally, the relevant parameter is the tensor-to-scalar ratio $r$. This is a good observable for testing large-field inflation and some models such as chaotic inflation \cite{Linde:1983gd} with a monomial potential with power larger or equal to two are strongly disfavoured \cite{Planck:2018jri}. There is no such luxury for small-field inflation models though. Small-field usually corresponds to a smaller energy scale of inflation. This amounts to a smaller tensor-to-scalar ratio. One representative type of small-field inflation model is hybrid inflation \cite{Linde:1993cn}. Basically, it works like a phase transition without temperature. Because the waterfall field (whose potential provides the vacuum energy density during inflation) in this model has a Higg-like potential form, hybrid inflation is often connected with particle physics such as grand unified theories (GUT), supersymmetry (SUSY), or supergravity (SUGRA). In this work, we consider a hilltop version of sneutrino hybrid inflation \cite{Antusch:2004hd, Antusch:2010mv}. This model may be connected to SUSY GUT since the right-handed neutrino belongs to the \textbf{16} representation of SO(10) grand unification. We also briefly consider nonthermal leptogenesis after inflation.

Baryogenesis is an interesting topic that connects particle physics and cosmology. The purpose is to answer the question about the asymmetry between matter and antimatter, namely why our universe is made out of one of them instead of the other. Among the theories of baryogenesis, we will consider leptogenesis \cite{Fukugita:1986hr} after inflation. 
Neutrino oscillations \cite{Super-Kamiokande:1998kpq} provide a concrete evidence for neutrino mass. The small neutrino masses can be elegantly explained by the seesaw mechanism \cite{Yanagida:1980xy, Gell-Mann:1979vob} where a right-handed neutrino with heavy Majorana mass is responsible for light neutrino mass.
The idea is to convert the lepton asymmetry generated by the decay of the heavy right-handed neutrino into the baryon asymmetry by the sphaleron \cite{Kuzmin:1985mm} process. If supersymmetry (SUSY) is considered\footnote{One benefit (among many others) to consider SUSY is that radiative correction to the Higgs mass via right-handed neutrino can be stabilized.}, one can have leptogenesis via the decay of sneutrino condensation \cite{Murayama:1993em}.
In the model where the sneutrino is the inflaton field, there is a ready-made sneutrino condensation.

\section{inflation}
The simplest inflation models have a single scalar field $\psi$ called the inflaton field with the potential $V(\psi)$. The equation of motion of a homogeneous scalar field $\psi$ in an expanding universe is
\begin{equation}
\ddot{\psi}+3H\dot{\psi}+V^\prime=0,
\end{equation}
where $H$ is the Hubble parameter.
This usually cannot be solved analytically and a slow-roll approximation scheme is considered.
The equation of motion is approximated as
\begin{equation}
3H\dot{\psi}+V^\prime=0.
\label{sreq}
\end{equation}
The slow-roll parameters are defined by (for example, see \cite{Dimopoulos:2022wzo} for a textbook review)
\begin{equation}
\epsilon \equiv \frac{M_P^2}{2}\left(\frac{V^\prime}{V} \right)^2,   \;\;\;   \eta \equiv M_P^2 \frac{V^{\prime\prime}}{V}.
\label{slow}
\end{equation}
The spectrum is
\begin{equation}
P_R=\frac{1}{12\pi^2 M_P^6}\frac{V^3}{V^{\prime 2}}.
\end{equation}
The spectral index can be expressed in terms of the slow-roll parameters as
\begin{equation}
n_s\equiv 1+\frac{d\ln P_R}{d\ln k}=1+2\eta-6\epsilon,
\label{index}
\end{equation}
where $k$ is the comoving wave number for the perturbation.
Current observation from cosmic microwave radiation (CMB) gives $n_s=0.9649 \pm 0.0042$ at 68\% CL \cite{Planck:2018jri}. We will simply choose $n_s=0.96$ in the following discussion. 
One important point here is that the spectrum is red, namely $n_s<1$. This means the amplitude of the spectrum is larger for longer wavelength perturbations as can be seen from the definition of Eq.~(\ref{index}).
The tensor-to-scalar ratio is
\begin{equation}
r=16\epsilon.
\label{r}
\end{equation}
There is currently only an upper bound of $r$, which is something like $r < 0.1$ \cite{Planck:2018jri}. 
 
Under the slow-roll approximation, the inflaton field value in order to achieve $N$ e-folds of inflation $\psi(N)$ until the end of inflation at $\psi_{\mathrm{end}}$ is given by
\begin{equation}
N=\frac{1}{M_P^2}\int^{\psi(N)}_{\psi_{\mathrm{end}}}\frac{V}{V^\prime}d\psi.
\label{efold}
\end{equation} 
From observation, $P^{1/2}_R \simeq 5 \times 10^{-5}$ at $N \simeq 60$ \cite{Planck:2018jri}. We call this the CMB normalization. For slow-roll inflation with a single inflaton field, a model of inflation means a suitable form of $V(\psi)$. One of the main challenges to solving a model of inflation with your favorite potential $V(\psi)$ is to see if you can integrate Eq.~(\ref{efold}).

From Eqs.~(\ref{r}) and (\ref{efold}), one can find that small-field inflation implies small $\epsilon$ and hence small $r$. Therefore we can write the spectral index of Eq.~(\ref{index}) as
\begin{equation}
n_s=1+2\eta.
\label{indexh}
\end{equation}
A red spectrum implies $\eta<0$, and from Eq.~(\ref{slow}), this means a concave downward potential. This suggests a potential of the hilltop form.

\section{sneutrino hybrid inflation and type III hilltop inflation}

In this section, we discuss sneutrino hybrid inflation \cite{Antusch:2004hd, Antusch:2010mv} and its connection to hilltop inflation. Sneutrino hybrid inflation is a supersymmetric hybrid inflation.
Let us consider the superpotential
\begin{equation}
W=\kappa \hat{S}\left( \frac{\hat{\phi}^4}{{M^\prime}^2}-M^2 \right)+\frac{(\lambda_N)_{ij}}{M_\ast} \hat{N}^i \hat{N}^j \hat{\phi} \hat{\phi}+(y_\nu)_{ij}\hat{N}^i \hat{h}_a \epsilon^{ab}\hat{L}^j_b+\dots,
\label{sup}
\end{equation}
where $\hat{N}^i$ is the superfield which contains the sneutrino inflaton field $\tilde{N}$. The index $i=1,2,3$ denotes generations. We assume the inflaton field as the lightest right-handed sneutrino field $N^1$. This will be further applied in section \ref{lep} in the discussion of leptogenesis. Furthermore, $\hat{\phi}$ is the superfield that contains a scalar field $\phi$ called the waterfall field in the context of hybrid inflation, and $\hat{S}$ is the superfield which contains a singlet scalar field $S$. The field value of $S$ is assumed to be zero during inflation which is stabilized by Hubble-induced mass from SUGRA correction. The superfields $\hat{L}$ and $\hat{h}$ contain the standard model leptons and (up-type) Higgs.
 
During inflation, the sneutrino inflaton field $\tilde{N}$ is assumed to have a large field value which stabilizes the $\phi$ field at zero through the term with the coupling $(\lambda_N)_{ij}$ and stabilizes the scalar part of $\hat{h}$ and $\hat{L}^i$ at zero through the term with the Yukawa coupling $(y_\nu)_{ij}$. This result in a large vacuum energy $V_0=\kappa^2 M^4$ which drives inflation. 

The global supersymmetry F-term scalar potential is given by
\begin{equation}
V_{\mathrm{SUSY}}=\kappa^2 \left(  \frac{|\phi|^4}{{M^\prime}^2}-M^2\right)^2+\frac{4\lambda_N^2}{M_\ast^2}(|\tilde{N}|^4|\phi|^2+|\tilde{N}|^2|\phi|^4)
\label{glo}
\end{equation}
Because $\tilde{N}$, $\phi$, and $S$ are gauge singlets, there is no D-term potential in this model.

For SUGRA corrections, let us consider a K\"{a}hler potential for the bosonic components of the superfields
\begin{equation}
\begin{aligned}
K=&|S|^2+|\phi|^2+|\tilde{N}|^2+\kappa_S \frac{|S|^4}{4M_P^2}+\kappa_N \frac{|\tilde{N}|^4}{4M_P^2}+\kappa_\phi \frac{|\phi|^4}{4M_P^2}+  \\&  \kappa_{S\phi}\frac{|S|^2|\phi|^2}{M_P^2}+\kappa_{SN}\frac{|S|^2|\tilde{N}|^2}{M_P^2}+\kappa_{N\phi}\frac{|\tilde{N}|^2|\phi|^2}{M_P^2}+\kappa_{SNN}\frac{|S|^2|\tilde{N}|^4}{M_P^4}.
\end{aligned}
\end{equation}
This is the same as that considered in \cite{Antusch:2004hd} except for the last term. These terms with Planck mass appear because supergravity is a non-renormalizable theory and it is assumed that the cut-off scale is Planck mass.
Here we neglect higher-order terms because we are considering a small-field model.
We will also neglect radiative corrections which presumably are subdominant. 
The SUGRA F-term scalar potential is given by
\begin{equation}
V=e^{\frac{K}{M_P^2}}\left[ \left(W_m + \frac{WK_m}{M_P^2} \right)^\dagger K^{m^\dagger n}  \left(W_n + \frac{WK_n}{M_P^2} \right)-\frac{3|W|^2}{M_P^2}  \right],
\label{sugrapo}
\end{equation}
where $K^{m^\dagger n}$ is the inverse matrix of
\begin{equation}
K_{m^\dagger n}=\frac{\partial^2 K}{\partial \phi^\dagger_m \partial \phi_n}
\end{equation}
and $\phi_n$ denotes different scalar fields in the K\"{a}hler potential. 
Although the calculation of Eq.~(\ref{sugrapo}) may look complicated at first sight, it can be simplified during inflation since $S=\phi=W=0$. In this case, we only have to consider the term with
\begin{equation}
W_{S}=-\kappa M^2,
\end{equation}
and
\begin{equation}
K_{S^\dagger S}=1+\kappa_{SN}\frac{|\tilde{N}|^2}{M_P^2}+\kappa_{SNN}\frac{|\tilde{N}|^4}{M_P^4}.
\end{equation}
The inverse can be obtained as
\begin{equation}
K^{S^\dagger S}=\frac{1}{1+\kappa_{SN}\frac{|\tilde{N}|^2}{M_P^2}+\kappa_{SNN}\frac{|\tilde{N}|^4}{M_P^4}} \simeq 1-\kappa_{SN}\frac{|\tilde{N}|^2}{M_P^2}+\kappa_{SN}^2 \frac{|\tilde{N}|^4}{M_P^4}        -\kappa_{SNN}\frac{|\tilde{N}|^4}{M_P^4}.
\end{equation}
We could expand the exponential factor as 
\begin{equation}
e^{\frac{K}{M_P^2}} \simeq 1+\frac{|\tilde{N}|^2}{M_P^2}+\frac{|\tilde{N}|^4}{2M_P^4}+\kappa_N \frac{|\tilde{N}|^4}{4M_P^2}
\end{equation}
The scalar potential is given by\footnote{This result is a little bit different from that in \cite{Antusch:2004hd}.}
\begin{equation}
\begin{aligned}
V&=e^{\frac{K}{M_P^2}} K^{S^\dagger S} V_0 \\& \simeq V_0 \left[ 1+(1-\kappa_{SN})\frac{|\tilde{N}|^2}{M_P^2}+  \left( \frac{1}{2}+\kappa_{SN}^2-\kappa_{SN}-\kappa_{SNN}+\frac{1}{4}\kappa_N \right)\frac{|\tilde{N}|^4}{M_P^4}   \right],   
\end{aligned}
\label{gra}
\end{equation}
where $V_0=|W_S|^2=\kappa^2 M^4$.
We also have to calculate the SUGRA correction to the mass term for the waterfall field $\phi$. It is not so difficult because when $S=0$, the fields $\tilde{N}$ and $\phi$ are symmetric in the potential, therefore we can simply change $\kappa_{SN}$ to $\kappa_{S\phi}$. The quadratic terms for $\phi$ including that from Eq.~(\ref{glo}) is
\begin{equation}
\frac{4\lambda_N^2}{M_\ast^2}|\tilde{N}|^4|\phi|^2+(1-\kappa_{S\phi})V_0\frac{|\phi|^2}{M_P^2}.
\label{pmass}
\end{equation}
From now on, we consider the real fields $\tilde{N}_R=\sqrt{2}|\tilde{N}|$ and $\phi_R=\sqrt{2}|\phi|$. The scalar potential in Eq.~(\ref{gra}) becomes
\begin{equation}
\begin{aligned}
V&=V_0 \left[ 1+(1-\kappa_{SN})\frac{\tilde{N}_R^2}{2M_P^2}+  \left( \frac{1}{2}+\kappa_{SN}^2-\kappa_{SN}-\kappa_{SNN}+\frac{1}{4}\kappa_N \right)\frac{\tilde{N}_R^4}{4M_P^4}   \right] \\& \equiv V_0 \left(   1+\gamma   \frac{\tilde{N}_R^2}{2M_P^2}  +\delta \frac{\tilde{N}_R^4}{4M_P^4}  \right)
\\& \equiv V_0 \left( 1+ \frac{1}{2}\eta_0 \frac{\tilde{N}_R^2}{M_P^2}\right)-\lambda \tilde{N}_R^4.   
\label{po}
\end{aligned}
\end{equation}
From Eq.~(\ref{pmass}), the mass squared for the waterfall field $\phi_R$ is
\begin{equation}
m^2_{\phi_R}=\lambda^2_N \frac{\tilde{N}_R^4}{M_\ast^2}-\beta\frac{V_0}{M_P^2},
\label{mpr}
\end{equation}
where $\beta \equiv \kappa_{S\phi}-1$. Inflation ends when the waterfall field starts to become tachyonic. This happens when $\tilde{N}_R$ reaches a critical value 
\begin{equation}
\tilde{N}_{Rc}^2=\sqrt{\beta V_0}\frac{M_\ast}{\lambda_N M_P}.
\label{nc}
\end{equation}
It is required here that $\beta>0$. Inflation ends more abruptly with a larger $\beta$ (see Appendix \ref{ined}).
Eventually, $\phi_R$ rolls to the potential minimum and develops a vacuum expectation value
\begin{equation}
\langle \phi_R  \rangle=\sqrt{2M^\prime M}.
\label{pvev}
\end{equation}

The parameterization of $\gamma$ and $\delta$ in Eq.~(\ref{po}) is adopted from \cite{Antusch:2004hd} where it is assumed that $\delta=0$ and $\gamma$ can be either positive or negative. For example, it is found that the spectral index $n_s=1+2\gamma$, as can be seen from Eq.~(\ref{indexh}) by assuming $V \simeq V_0$. If we need $n_s=0.96$, then $\gamma=-0.02$. Unfortunately, if $\gamma$ is negative (with $\delta=0$), the inflaton field actually would roll toward the direction where the field value grows (opposite to the expected direction) and inflation will not end properly to arrive at the critical value given by Eq.~(\ref{nc}) and eventually goes to zero inflaton field value. This shows the important role played by the $\delta$ or $\lambda$ of the quartic term.

\begin{figure}[t]
  \centering
\includegraphics[width=0.6\textwidth]{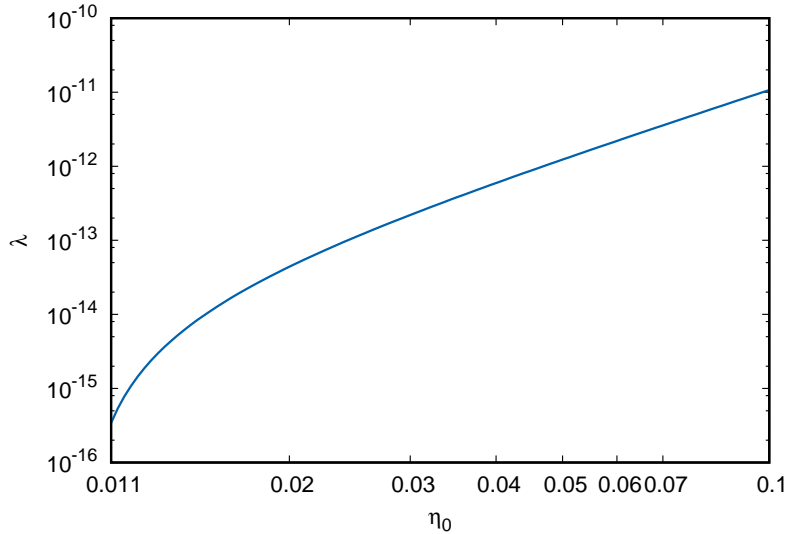}
  \caption{$\lambda$ as a function of $\eta_0$.}
  \label{fig001}
\end{figure}

The parameterization of $\eta_0$ and $\lambda$ in Eq.~(\ref{po}) with $\eta_0>0$ and $\lambda>0$ belong to a special case ($p=4$) of what we may call type III hilltop inflation \cite{Kohri:2007gq} (see Appendix \ref{hilltopi}). Interestingly, this model can be solved analytically under slow-roll approximation and assuming the $V_0$ term dominates. By imposing CMB normalization $P^{1/2}_R=5 \times 10^{-5}$ and $n_s=0.96$ we have \cite{Kohri:2013gva} (see Appendix \ref{hilltopi} for more details)
\begin{equation}
\lambda=1.1\times 10^{-8}(\eta_0+0.02)(\eta_0-0.01)^2.
\label{rela}
\end{equation}
We plot $\lambda$ as a function of $\eta_0$ in Fig \ref{fig001}. 
From Eq.~(\ref{po}), $\eta_0=1-\kappa_{SN}$. From dimensional analysis, the nature value of the parameter $\kappa_{SN}$ is about $\kappa_{SN}=\mathcal{O}(1)$. As can be seen in the plot, we need $0.01 \lesssim  \eta_0 \lesssim  0.1$. This may require a little bit of tuning, but certainly not fine-tuning. For example, $\eta_0=0.04$ can be achieved by $\kappa_{SN}=0.96$. On the other hand, from Eq.~(\ref{po})
\begin{equation}
\lambda=-\left( \frac{1}{2}+\kappa_{SN}^2-\kappa_{SN}-\kappa_{SNN}+\frac{1}{4}\kappa_N \right)\frac{V_0}{M_P^4}.
\end{equation}
The nature value of the parameters $\kappa_{SN}$, $\kappa_{SNN}$, and $\kappa_{N}$ are all of $\mathcal{O}(1)$ from dimensional analysis, therefore we just assume $\lambda=V_0/M_P^4$. In principle, these parameters can be calculated if we have a theory of quantum gravity. From Fig.~\ref{fig001}, we can see that the required $\lambda$ is quite small. Let us take $\lambda=10^{-12}$ as an example. This just tells us that the scale of inflation is $V_0^{1/4}=10^{-3}M_P =2.4\times 10^{15}\mbox{ GeV}$ which is about the GUT scale.

\section{nonthermal leptogenesis}
\label{lep}
After inflation, both the sneutrino inflaton field and the waterfall field start to oscillate and decay. For simplicity, we assume the waterfall field decays first and achieves its vacuum expectation value so that the reheating process is governed by the sneutrino field. The sneutrino field decays through the coupling $y_\nu$ to lepton and Higgsino or slepton and Higgs particles in Eq.~(\ref{sup}). From Eqs.~(\ref{glo}) and (\ref{pvev}), the mass of the lightest right-handed sneutrino (inflaton) is given by\footnote{The idea that a right-handed sneutrino acquires a mass from the vacuum expectation of the waterfall field was considered in \cite{Berezhiani:2001xx} where right-handed sneutrino is not the inflaton.} 
\begin{equation}
m_{N^1}=\frac{2(\lambda_N)_{11}MM^\prime}{M_\ast}.
\label{mn}
\end{equation}
The decay width is
\begin{equation}
\Gamma_{N^1}=\frac{(y^\dagger_\nu y_\nu)_{11}}{4\pi}m_{N^1}.
\end{equation}
The sneutrino field decays when $\Gamma_{N^1} \sim H$ at the energy density
\begin{equation}
\rho \sim T_R^4 \sim \Gamma_{N^1}^2 M_P^2.
\label{crie}
\end{equation}
Therefore the reheating temperature is
\begin{equation}
T_R \sim \sqrt{\Gamma_{N^1}M_P}.
\end{equation}

Nonthermal leptogenesis means $m_{N^1}  \gg T_R$, namely $N^1$ is not generated in the thermal bath. In this case, the baryon asymmetry is
\begin{equation}
\frac{n_B}{n_\gamma} \sim \epsilon_1 \frac{T_R}{m_{N^1}},
\label{bar}
\end{equation}
where $\epsilon_1$ is the lepton asymmetry parameter. It is defined as
\begin{equation}
\epsilon_1 \equiv \frac{\Gamma(\tilde{N}^1\rightarrow L+H_u)-\Gamma(\tilde{N}^1\rightarrow \overline{L}+\overline{H_u})}{\Gamma(\tilde{N}^1\rightarrow L+H_u)+\Gamma(\tilde{N}^1\rightarrow \overline{L}+\overline{H_u})}.
\end{equation}
The observed baryon-to-photon ratio (with standard deviation) is \cite{Fields:2019pfx}
\begin{equation}
\frac{n_B}{n_\gamma}=(6.104 \pm 0.058)\times 10^{-10}.
\end{equation}
There is an upper bound of $\epsilon_1$ for a hierarchical spectrum of right-handed neutrinos given by \cite{Hamaguchi:2001gw, Davidson:2002qv}
\begin{equation}
\epsilon_1 < \frac{3}{8\pi}\frac{\sqrt{\Delta m^2_{\mathrm{atm}}}m_{N^1}}{\langle v \rangle^2},
\label{atm}
\end{equation}
where $\langle v \rangle \simeq 174 \mbox{ GeV}$ is the vacuum expectation value of the up-type Higgs and $\Delta m_{31}^2 \simeq 2.6 \times 10^{-3} \mbox{ eV}^2$ is the atmospheric neutrino mass squared difference. By using Eqs.~(\ref{bar}) and (\ref{atm}), a simple estimation for the condition of successful leptogenesis is just
\begin{equation}
T_R \gtrsim 10^6\mbox{ GeV}.
\end{equation}
For example, for $m_{N^1}=10^8\mbox{ GeV}$ and $(y^\dagger_\nu y_\nu)_{11}=10^{-12}$, we have $T_R \sim 10^6 \mbox{ GeV}$.

\section{Conclusion}
\label{con}

In this work, we realize a concrete model of type III hilltop inflation in the framework of sneutrino hybrid inflation. We neglect the subdominant contribution of radiative corrections to the potential which may be considered by the method of \cite{Lin:2008ys}, but the calculation will be more involved. 

We have shown that the model produces a perfect fit for the spectral index $n_s$ with a negligible tensor-to-scalar ratio $r$ which also fits the current experimental constraint. In this model, the inflaton is the lightest right-handed sneutrino field. This model potentially may be embedded into grand unified theories since the existence of right-handed sneutrino is predicted in GUT such as SO(10).

Sneutrino hybrid inflation can be generalized into a class of models called tribrid inflation \cite{Antusch:2009vg, Antusch:2015tha, Masoud:2021prr}. In principle, type III hilltop inflation may also be applicable in (some of) these cases. We leave the further investigation to future works.

\appendix

\section{the end of inflation}
\label{ined}
In this section, we estimate the condition for sneutrino hybrid inflation to end abruptly when the waterfall field becomes tachyonic.
Close to the end of inflation, the quadratic term dominates the quartic term. The potential is
\begin{equation}
V=V_0+\frac{\eta_0 V_0}{2}\frac{\tilde{N}^2_R}{M_P^2}.
\end{equation}
Suppose slow-roll inflation continues after $\tilde{N}_R$ reaches $\tilde{N}_{Rc}$, the equation of motion is given by Eq.~(\ref{sreq}) as
\begin{equation}
3H\dot{\tilde{N}}_R=-V^\prime \simeq -\frac{\eta_0 V_0}{M_P^2}\tilde{N}_{Rc}
\end{equation}
During one Hubble time $\Delta t=1/H$, the variation of $\tilde{N}_R$ is
\begin{equation}
\Delta \tilde{N}_R=\eta_0 \tilde{N}_{Rc},
\end{equation}
where we have used $V_0=3H^2 M_P^2$.
From Eqs.~(\ref{mpr}) and (\ref{nc}) we obtain
\begin{equation}
m^2_{\phi_R}=3H^2 \beta \left( (1-\eta_0)^4-1 \right).
\end{equation}
For example, if $\eta_0=0.04$, $m^2_{\phi_R}=-0.45 \beta H^2$. We need $\beta>2.2$ for the slow-roll of the waterfall field $\phi_R$ to fail within one Hubble time. However, this is not a very crucial point because even if inflation lasts a little bit longer, our conclusion does not change due to some allowed uncertainty of the number of e-folds at the horizon exit. 

\section{hilltop inflation}
\label{hilltopi}
In this section, we recall the hilltop inflation models proposed in \cite{Kohri:2007gq}.
For an inflation model with a potential of the form,
\begin{equation}
V(\psi)=V_0\left( 1+\frac{1}{2}\eta_0\frac{\psi^2}{M_P^2} \right)-\lambda \frac{\psi^p}{M_P^{p-4}},
\end{equation}
There are three types of hilltop inflation models from the above parameterization. In all three cases, $\lambda>0$. 
\begin{itemize}
\item Type I, $\eta_0 \leq 0$ and $p>2$. 
\item Type II, $\eta<0$ and $p<0$.
\item Type III, $\eta>0$ and $p>2$.
\end{itemize}
By assuming $V \simeq V_0$, Eq.~(\ref{efold}) can be integrated analytically to obtain
\begin{eqnarray}
\left( \frac{\psi}{M_P} \right)^{p-2}&=&\left( \frac{V_0}{M_P^4}\right)\frac{\eta_0e^{(p-2)N\eta_0}}{\eta_0 x+p\lambda(e^{(p-2)N\eta_0}-1)},\\
x&\equiv&\left( \frac{V_0}{M_P^4} \right)\left( \frac{M_P}{\psi_{\mathrm{end}}} \right)^{p-2}.
\end{eqnarray}
The spectrum and the spectral index are 
\begin{eqnarray}
P_R&=&\frac{1}{12\pi^2}\left( \frac{V_0}{M_P^4} \right)^{\frac{p-4}{p-2}}e^{-2N\eta_0}\frac{[p\lambda(e^{(p-2)N\eta_0}-1)+\eta_0 x]^{\frac{2p-2}{p-2}}}{\eta_0^{\frac{2p-2}{p-2}}(\eta_0 x-p\lambda)^2},  \\
n_s&=&1+2\eta_0 \left[1-\frac{\lambda p(p-1)e^{(p-2)N\eta_0}}{\eta_0 x+p\lambda(e^{(p-2)N\eta_0}-1)} \right].
\end{eqnarray}
From the above equations, we have \cite{Kohri:2013gva}
\begin{equation}
\lambda=\frac{(12\pi^2 P_R)^{\frac{p-2}{2}}}{p[2(p-1)]^{p-1}}\left( \frac{V_0}{M_P^4} \right)^{-\frac{p-4}{2}}(2\eta_0-n_s+1)(2(p-2)\eta_0+n_s-1)^{p-2}.
\end{equation}
Eq.~(\ref{rela}) corresponds to the case $p=4$ of Type III.

\acknowledgments
This work is supported by the National Science and Technology Council (NSTC) of Taiwan under Grant No. NSTC 111-2112-M-167-002.

\end{document}